\documentclass[doublecol]{epl2}
\usepackage{amssymb,amsmath,graphicx,setspace,color,rotating,subfigure,url}
\usepackage{extarrows}
\title{Weiqi games as a tree: Zipf's law of openings and beyond}

\author{Li-Gong Xu\inst{1} \and Ming-Xia Li\inst{1,2,3} \and Wei-Xing Zhou\inst{3,4,5}\footnote{e-mail: wxzhou@ecust.edu.cn}}
\shortauthor{L.-G. Xu \etal}

\institute{
  \inst{1} School of Sports Science and Engineering, East China University of Science and Technology, Shanghai 200237, China\\
  \inst{2} Postdoctoral Research Station, East China University of Science and Technology, Shanghai 200237, China\\
  \inst{3} Research Center for Econophysics, East China University of Science and Technology, Shanghai 200237, China\\
  \inst{4} School of Business, East China University of Science and Technology, Shanghai 200237, China\\
  \inst{5} Department of Mathematics, East China University of Science and Technology, Shanghai 200237, China\\
}


\abstract{
 Weiqi is one of the most complex board games played by two persons. The placement strategies adopted by Weiqi players are often used to analog the philosophy of human wars. Contrary to the western chess, Weiqi games are less studied by academics partially because Weiqi is popular only in East Asia, especially in China, Japan and Korea. Here, we propose to construct a directed tree using a database of extensive Weiqi games and perform a quantitative analysis of the Weiqi tree. We find that the popularity distribution of Weiqi openings with a same number of moves is distributed according to a power law and the tail exponent increases with the number of moves. Intriguingly, the superposition of the popularity distributions of Weiqi openings with the number of moves no more than a given number also has a power-law tail in which the tail exponent increases with the number of moves, and the superposed distribution approaches to the Zipf law. These findings are the same as for chess and support the conjecture that the popularity distribution of board game openings follows the Zipf law with a universal exponent. We also find that the distribution of out-degrees has a power-law form, the distribution of branching ratios has a very complicated pattern, and the distribution of uniqueness scores defined by the path lengths from the root vertex to the leaf vertices exhibits a unimodal shape. Our work provides a promising direction for the study of the decision making process of Weiqi playing from the angle of directed branching tree.
}

\begin{document}

\maketitle


\section{Introduction}
\label{S1:Intro}

Weiqi, also called Go in Japan and Budak in Korea, originated over 3000 years ago in China and is probably the oldest board game in the world. The standard modern Weiqi game is played on a $19\times19$ board. It is played by two players using black and white stones in turn. The rules of Weiqi playing are very simple. However, its strategies are very complex, much more than western chess. Specifically, Weiqi is PSPACE-hard \cite{Lichtenstein-Sipser-1980-JACM} and EXPTIME-complete \cite{Robson-1983-CONF} and certain subproblems of the game are PSPACE-complete \cite{Crasmaru-Tromp-2002-LNCS}. It has been shown that the state-space complexity of Weiqi is $\sim10^{171}$ and its game-tree complexity is $\sim10^{360}$ \cite{Allis-1994-PhD,Tromp-Farneback-2007-LNCS}. For comparison, these two numbers are $10^{47}$ and $10^{123}$ for chess.

Although computer programs are able to defeat top-tier professional players for many mind board games including chess and Chinese chess (Xiangqi), the situation is completely different for Weiqi. It is even very hard for a computer to judge if a cluster of stones is alive or dead, and it is much harder to solve correctly a lot of life-and-death problems \cite{Benson-1976-IS,Chen-Chen-1999-IS}. Certainly, there are progresses on predicting life or death in the game of Weiqi through learning examples extracted from game records \cite{VanderWerf-Winands-VanDenHerik-Uiterwijk-2005-IS}. No doubt, the game of Weiqi has been a big challenge to researchers in the field of Artificial Intelligence (AI) \cite{Thorp-Walden-1972-IS,Bouzy-Cazenave-2001-AI,Muller-2002-AI,Oshima-Yamada-Endoa-2011-PCS}. Interestingly, in modern neuroscience, it has been shown that long-term trained
Weiqi players ``developed larger regions of white matter with increased fractional anisotropy values in the frontal, cingulum, and striato-thalamic areas that are related to attentional control, working memory, executive regulation, and problem-solving'' \cite{Lee-Park-Jung-Kim-Oh-Choi-Jang-Kang-Kwon-Ni-2010-Ni}.

Recently, there are also efforts to study Weiqi games from the perspective of statistical physics. Liu, Dou and Lu defined an area, say a $5\times5$ plaquette, centered by the current move and obtained a contextual pattern with successive moves within the plaquette. It is found that the occurrence distribution of contextual patterns follows Zipf's law \cite{Liu-Dou-Lu-2008-LNCS}. Harre et al. investigated the probabilities of different moves within the $7\times7$ corner plaquette to quantify the lack of predictability of experts and how this changes with their levels of skill \cite{Harre-Bossomaier-Gillet-Snyder-2011-EPJB}. Georgeot and Giraud defined the 1107 plaquettes of size $3\times3$ with empty centers as vertices and, if the distance between two successive moves is smaller than a preset value, the two associated vertices are connected with an arrow \cite{Georgeot-Giraud-2012-EPL}. Alternatively, when the atari status of the four nearest neighbor points from the center is taken into consideration, one obtains 2051 legal nonequivalent plaquettes or vertices with empty centers \cite{Huang-Coulom-Lin-2011-LNCS,Kandiah-Georgeot-Giraud-2014-EPJB}, and considering diamond-shape plaquettes by expanding the $3\times3$ plaquettes and four additional points results in 193 995 nonequivalent plaquettes with empty centers \cite{Kandiah-Georgeot-Giraud-2014-EPJB}. Very intriguing results have been obtained for these networks \cite{Harre-Bossomaier-Gillet-Snyder-2011-EPJB,Huang-Coulom-Lin-2011-LNCS,Georgeot-Giraud-2012-EPL,Kandiah-Georgeot-Giraud-2014-EPJB}.

In this Letter, we construct a directed Weiqi tree using a database of professional games,  which is a subset of the whole strategy tree studied in the AI literature. The Weiqi tree constructed here adopts the same method as in Ref.~\cite{Blasius-Tonjes-2009-PRL} for chess. The method takes the move order into consideration so that the network is a tree and the in-degrees of all non-root vertices are 1. We will study the distributions of the popularity of Weiqi openings, the out-degrees, the branching ratios and the uniqueness scores of the Weiqi tree.


\section{Data description}

The database of Weiqi games used in this Letter, called All-in-One Library of Weiqi Games (or ``x\'{u}n p\v{u} m\v{u} b\v{a}n'' in Chinese spell), is collected and maintained by the Gochess web site since more than 10 years ago, and can be downloaded from http://www.gochess.cn/. All the games were played by professionals. We use version 7.6 here, which has 69 031 games. After removing duplicate games, handicap games, transposon games, games on $13\times13$ board, life or dead problems, non-human played games, and games with less than 30 moves, there are 57 090 games left for further analysis.

\section{Construction of the Weiqi tree}

\begin{figure}[htb]
  \centering
  \includegraphics[width=0.9\linewidth]{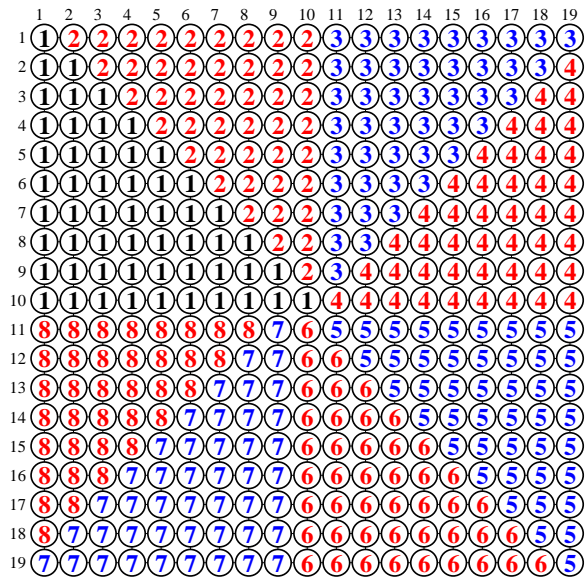}
  \caption{\label{Fig:Weiqi:Zones} Partitioning the Weiqi board into eight zones: $Z_1=\{x\leq{y}\leq10\}$, $Z_2=\{y<x\leq10\}$, $Z_3=\{x+y\leq20, x>10\}$,$Z_4=\{x+y>20, y\leq10\}$, $Z_5=\{10<y\leq{x}\}$, $Z_6=\{x<y, x\geq10\}$, $Z_7=\{20\leq{x+y}, x<10\}$, and $Z_8=\{y+x<20, x<10, y>10\}$. The crosses in the first zone are determined. For other zones, it is optional to include the crosses at $x=y$, $x=20-y$, $x=10$ or $y=10$. Two zones both with even or odd numbers have the possibility to overlap after rotation, while two zones with even and odd numbers may overlap after flip and rotation.}
\end{figure}

We denote each game by ${\cal{G}}_i$. For a Weiqi game ${\cal{G}}_i$, the $k$-th move can be determined by a point ${\cal{M}}_k=(x_k,y_k)$ on the board, and the opening with $m$ moves is a sequence of $m$ points: ${\cal{G}}_{i,m}=\{{\mathcal{M}}_k: k=1,2,\cdots,m\}$. The coordinates $(x,y)$ of the $19\times19$ board are shown in fig.~\ref{Fig:Weiqi:Zones}.

Two main features of Weiqi games are rotation equivalence and flip equivalence, which means that two openings ${\cal{G}}_{i,m}$ and ${\cal{G}}_{j,m}$ overlap after certain rotation and/or flip. Equivalent rotations contain $90^\circ$, $180^\circ$ and $270^\circ$ clockwise rotations. Mathematically, we have
\begin{equation}
  {\cal{G}}_{i,m}(x, y) \equiv
  \left\{
  \begin{array}{lll}
    {\cal{G}}_{i,m}(20-y,x),     &  90^\circ\\
    {\cal{G}}_{i,m}(20-x, 20-y), & 180^\circ\\
    {\cal{G}}_{i,m}(y, 20-x),    & 270^\circ
  \end{array}
  \right.
  \label{Eq:Weiqi:Rotation}
\end{equation}
Equivalent flips can be conducted with respect to the $x$ or $y$ axis. Using $y$-flip, we have
\begin{equation}
  {\cal{G}}_{i,m}(x, y) \equiv   {\cal{G}}_{i,m}(20-x,y).
  \label{Eq:Weiqi:Flip}
\end{equation}
Combining a rotation and a flip, there are three more Weiqi patterns that are equivalent:
${\cal{G}}_{i,m}(20-y, 20-x)$, ${\cal{G}}_{i,m}(x, 20-y)$, and ${\cal{G}}_{i,m}(y, x)$.

As shown in fig.~\ref{Fig:Weiqi:Zones}, two openings ${\cal{G}}_{i,m}$ and ${\cal{G}}_{j,m}$ can be identical after proper rotation and slip. Hence, we preprocess the games to make them unique. We require that the first move locate in Zone 1 after suitable rotation and flip. If the first move is in Zone 1, i.e., ${\cal{M}}_1\in{Z_1}$, we do nothing. If ${\cal{M}}_1\in{Z_2}$, we flip it along $x=y$ (equivalently, rotate 90$^\circ$ clockwise and then flip along $x=10$), obtaining ${\cal{G}}_i(y,x)$ from ${\cal{G}}_i(x,y)$ for all games. If ${\cal{M}}_1\in{Z_3}$, we obtain ${\cal{G}}_i(y,20-x)$ through rotating ${\cal{G}}_i(x,y)$ by $90^\circ$ counterclockwise. If ${\cal{M}}_1\in{Z_4}$, we obtain ${\cal{G}}_i(20-x,y)$ through flipping ${\cal{G}}_i(x,y)$ along $x=10$. If ${\cal{M}}_1\in{Z_5}$, we obtain ${\cal{G}}_i(20-x,20-y)$ through rotating ${\cal{G}}_i(x,y)$ by $180^\circ$ counterclockwise. If ${\cal{M}}_1\in{Z_6}$, we obtain ${\cal{G}}_i(20-y,20-x)$ through flipping ${\cal{G}}_i(x,y)$ along $x+y=20$. If ${\cal{M}}_1\in{Z_7}$, we obtain ${\cal{G}}_i(20-y,x)$ through rotating ${\cal{G}}_i(x,y)$ by $270^\circ$ counterclockwise. If ${\cal{M}}_1\in{Z_8}$, we obtain ${\cal{G}}_i(x,20-y)$ through flipping ${\cal{G}}_i(x,y)$ along $y=10$. We stress that the transformation of coordinates is carried out for all moves, not only for the first move.

After the first step of transformation, the first move ${\cal{M}}_1$ has been placed in Zone 1. If $x_1=y_1$ and $x_2>y_1$, we flip the board along the diagonal $y=x$ so that ${\cal{G}}_i(x,y)\Rightarrow{\cal{G}}_i(y,x)$. Moreover, if $x_1=y_1$, $x_2=y_2$, $x_3>y_3$, we also flip the board along the diagonal $y=x$ so that ${\cal{G}}_i(x,y)\Rightarrow{\cal{G}}_i(y,x)$. More generally, if the first $k$ moves are on the diagonal and the $k+1$ move has $x_{k+1}>y_{k+1}$, we flip the board along $x=y$. Usually, Weiqi games with $k\geq3$ first moves on the diagonal are very rare.

\begin{figure}[htb]
  \centering
  \includegraphics[width=0.9\linewidth]{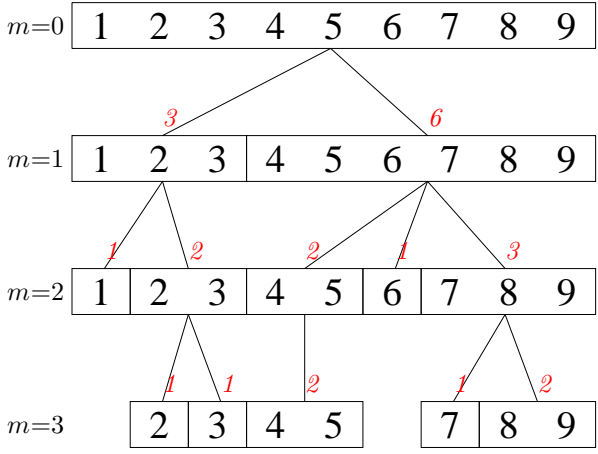}
  \caption{\label{Fig:Weiqi:Tree} An illustrative Weiqi tree with nine Weiqi games. The root vertex $V_{0,1}=\{{\cal{G}}_{i,0}: i=1,\cdots,9\}$ is the empty Weiqi board without any moves. After one move, at $m=1$, there are two vertices, $V_{1,1}=\{{\cal{G}}_{1,1}, {\cal{G}}_{2,1}, {\cal{G}}_{3,1}\}$ and $V_{1,2}=\{{\cal{G}}_{4,1}, {\cal{G}}_{5,1},{\cal{G}}_{6,1}, {\cal{G}}_{7,1}, {\cal{G}}_{8,1}, {\cal{G}}_{9,1}\}$. After two moves at $m=2$, $V_{1,1}$ generates two offspring vertices $V_{2,1}=\{{\cal{G}}_{1,2}\}$ and $V_{2,2}=\{{\cal{G}}_{2,2},{\cal{G}}_{3,2}\}$, while $V_{1,2}$ generates three offspring vertices $V_{2,3}=\{{\cal{G}}_{4,2},{\cal{G}}_{5,2}\}$, $V_{2,4}=\{{\cal{G}}_{6,2}\}$ and $V_{2,5}=\{{\cal{G}}_{7,2},{\cal{G}}_{8,2},{\cal{G}}_{9,2}\}$. Vertices $V_{2,1}$ and $V_{2,4}$ are leaf vertices because they contain only one game.}
\end{figure}

After the above transformations, all the openings are unique and standard, that is, any two openings ${\cal{G}}_{i,m}$ and ${\cal{G}}_{j,m}$ cannot overlap after further rotations and/or flips. Certainly, they might overlap without further rotations and flips. We can thus construct the Weiqi tree. Figure~\ref{Fig:Weiqi:Tree} illustrates the first four layers of the Weiqi tree with nine imaginary Weiqi games $\{{\cal{G}}_{i,0}: i=1,\cdots,9\}$. In this Weiqi tree, the first layer is the root vertex $V_{0,1}$, which contains all nine games and corresponds to the empty Weiqi board without any moves. We denote the $n$-th vertices on the $m$ layer as $\{V_{m,i}: i=1,2,\cdots,n\}$. The root vertex $V_{0,1}=\{{\cal{G}}_{i,0}: i=1,\cdots,9\}$ is the empty Weiqi board without any moves. After one move, at $m=1$, there are two vertices, $V_{1,1}=\{{\cal{G}}_{1,1}, {\cal{G}}_{2,1}, {\cal{G}}_{3,1}\}$ and $V_{1,2}=\{{\cal{G}}_{4,1}, {\cal{G}}_{5,1},{\cal{G}}_{6,1}, {\cal{G}}_{7,1}, {\cal{G}}_{8,1}, {\cal{G}}_{9,1}\}$. After two moves at $m=2$, $V_{1,1}$ generates two offspring vertices $V_{2,1}=\{{\cal{G}}_{1,2}\}$ and $V_{2,2}=\{{\cal{G}}_{2,2},{\cal{G}}_{3,2}\}$, while $V_{1,2}$ generates three offspring vertices $V_{2,3}=\{{\cal{G}}_{4,2},{\cal{G}}_{5,2}\}$, $V_{2,4}=\{{\cal{G}}_{6,2}\}$ and $V_{2,5}=\{{\cal{G}}_{7,2},{\cal{G}}_{8,2},{\cal{G}}_{9,2}\}$. Vertices $V_{2,1}$ and $V_{2,4}$ are leaf vertices because they contain only one game. Further, $V_{2,2}$ splits to $V_{3,1}=\{{\cal{G}}_{2,3}\}$ and $V_{3,2}=\{{\cal{G}}_{3,3}\}$, $V_{2,3}$ keeps unchanged to $V_{3,3}=\{{\cal{G}}_{4,3},{\cal{G}}_{5,3}\}$, and $V_{2,5}$ splits to $V_{3,4}=\{{\cal{G}}_{7,3}\}$ and $V_{3,5}=\{{\cal{G}}_{8,3},{\cal{G}}_{9,3}\}$. Each leaf vertex corresponds only to one game and cannot split. On the other hand, a vertex corresponding only to one game is always a leaf vertex.

Using the procedure described for the tree in fig.~\ref{Fig:Weiqi:Tree}, we construct the Weiqi tree for all the 57 090 games. The number of vertices increases from 1 at $m=0$ to 11428 at $m=9$, and then decreases to 518 at $m=30$. By definition, the final tree has 57 090 leaf vertices. We can define the out-degree of each vertex $V_{m,j}$, the popularity of each opening ${\cal{G}}_{i,m}$, the local branching of each vertex, the uniqueness of each game, and so on. We will investigate the properties of these tree features.

\section{Popularity of Weiqi openings}

The popularity $s_{m,j}$ of an opening ${\cal{G}}_{i,m}$ with $m$ moves, also know as weight \cite{Blasius-Tonjes-2009-PRL}, is the strength of node $V_{m,j}$ with ${\cal{G}}_{i,m}\in V_{m,j}$:
\begin{equation}
  s_{m,j} \triangleq s({\cal{G}}_{i,m}) \triangleq s(V_{m,j}) = |V_{m,j}|,
  \label{Eq:Weiqi:Popularity}
\end{equation}
where $|V|$ is the cardinality of set $V$. The relative popularity of an opening ${\cal{G}}_{i,m}$ is the ratio of the popularity $s_{m,j}$ to the total number $N$ of all games:
\begin{equation}
  w_{i,m} = s_{m,j}/N.
  \label{Eq:Weiqi:RelativePopularity}
\end{equation}
For the tree in fig.~\ref{Fig:Weiqi:Tree}, $s_{1,1}=3$ and $s_{1,2}= 6$ for $m=1$, $s_{2,1}=1$, $s_{2,2}= 2$, $s_{2,3}= 2$, $s_{2,4}= 1$, and $s_{2,5}= 3$ for $m=2$, and $s_{3,1}=1$, $s_{3,2}=1$, $s_{3,3}=2$, $s_{3,4}=1$ and $s_{3,5}= 2$ for $m=3$.

For the real tree, the first move was placed in 22 different crosses. The most popular move is ${\cal{G}}_{m=1}=[4,4]$ with $s([4,4])=39310$ and $w([4,4])=0.5930$, the second most popular move is ${\cal{G}}_{m=1}=[3,4]$ with $s([3,4])=25350$ and $w([3,4])=0.3824$, the third most popular move is ${\cal{G}}_{m=1}=[3,5]$ with $s([3,5])=743$ and $w([3,5])=0.0112$, the forth most popular move is ${\cal{G}}_{m=1}=[3,3]$ with $s([3,3])=389$ and $w([3,3])=0.0060$, and the fifth most popular move is ${\cal{G}}_{m=1}=[4,5]$ with $s([4,5])=243$ and $w([4,5])=0.0037$. Tianyuan, the central star, ranks the sixth with $s([10,10])=84$ and $w([4,5])=0.0013$. Other openings occur much less frequently, in which four openings have 3 games ($s({\cal{M}})=3$) and other 7 openings appear in only one game ($s({\cal{M}})=1$).

\begin{figure}[tb]
  \centering
  \includegraphics[width=0.9\linewidth]{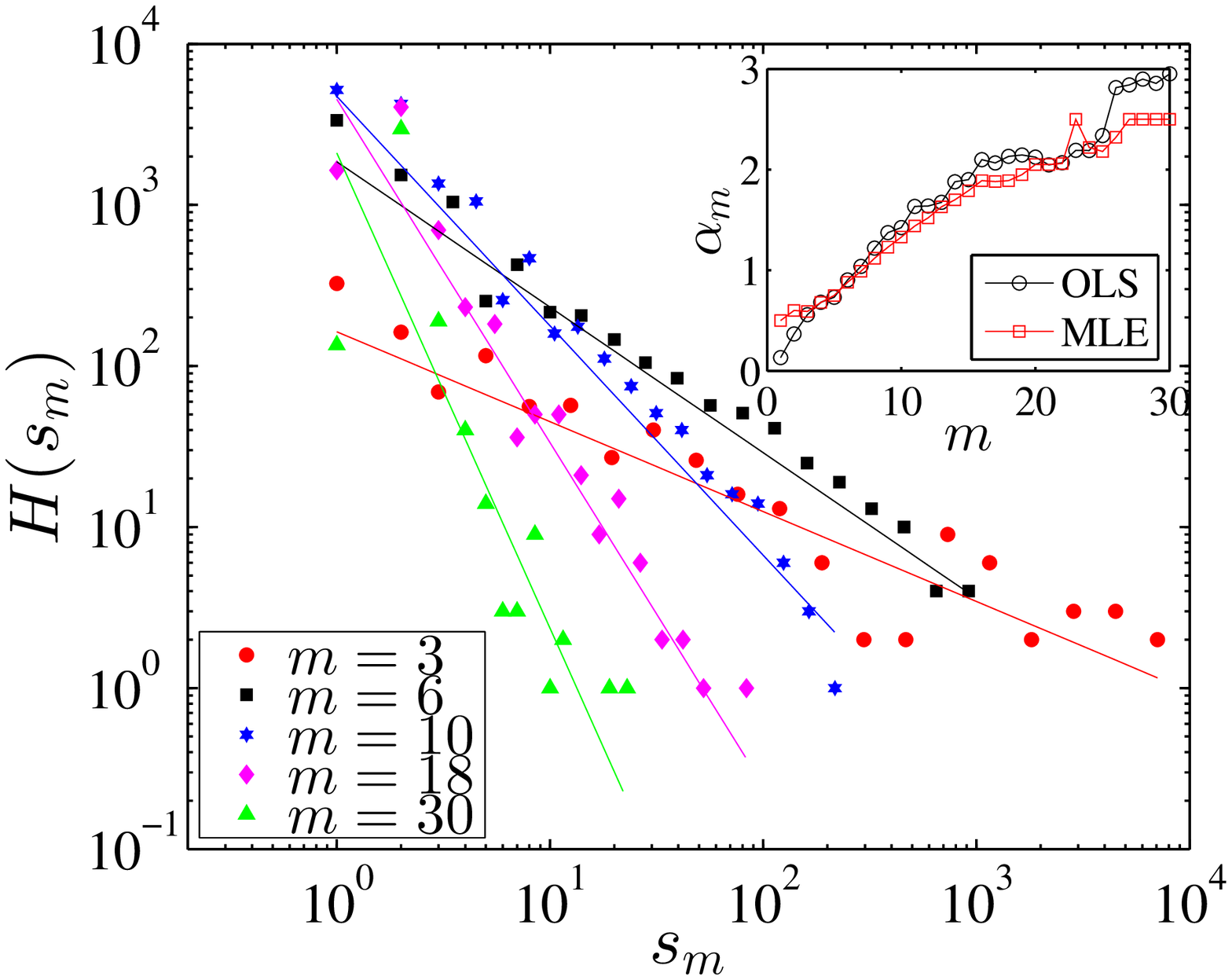}
  \includegraphics[width=0.9\linewidth]{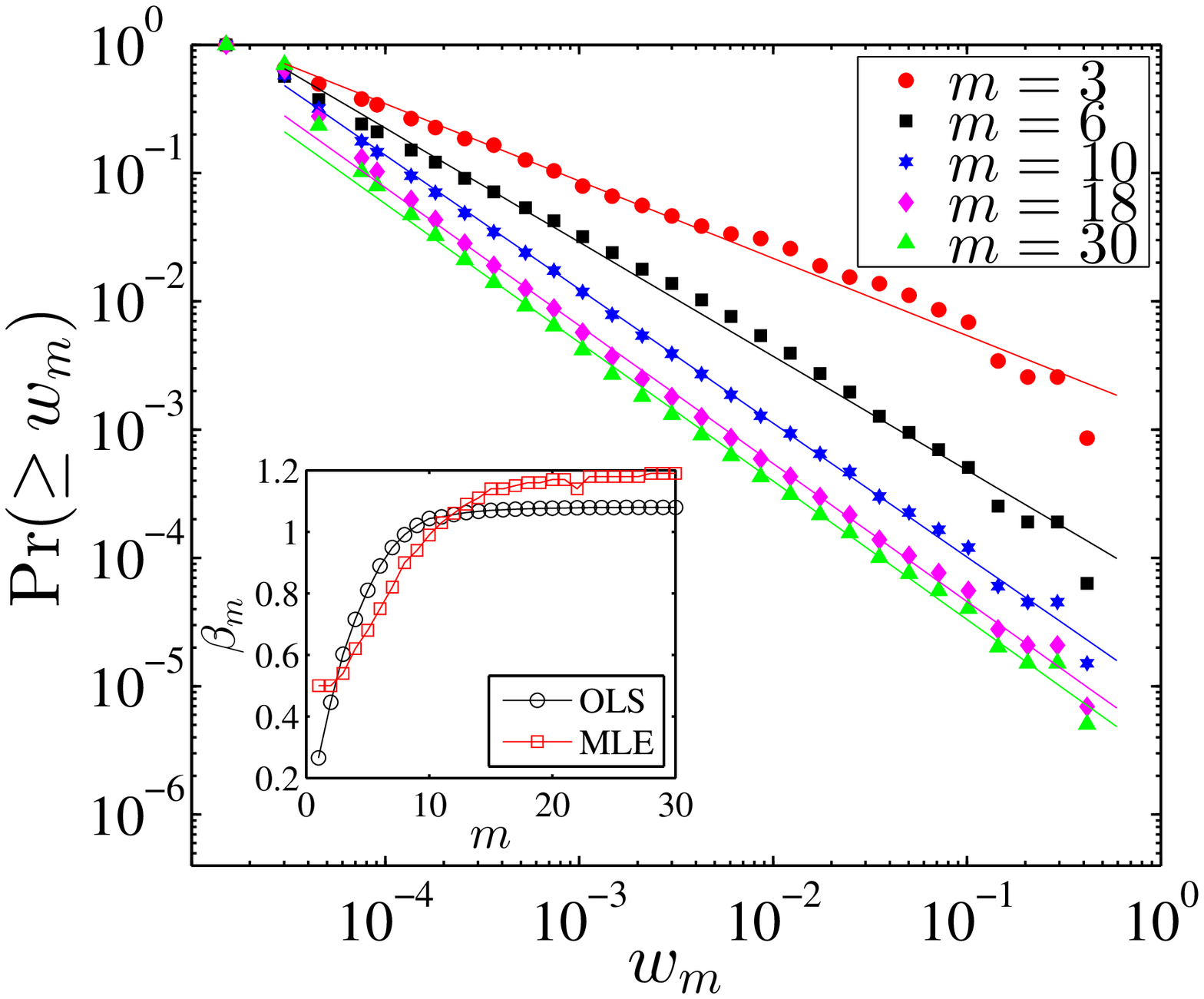}
  \caption{\label{Fig:Weiqi:Open:Popularity} (color online). (a) Histogram of popularity $s_m$ of openings ${\cal{G}}_m$ for $m=3$, 6, 10, 18, and 30. The five curves are histograms obtained with logarithmic binning. The solid lines are the best power-law fits. Inset: The slopes $\alpha_m$ of the regression lines obtained by the ordinary linear-squares regression and the maximum likelihood estimation as a function of $m$ for $m=1, \cdots, 30$. (b) Popularity distributions $\Pr(\geq{w_m})$ of pooled openings from ${\cal{G}}_1$ to ${\cal{G}}_m$ using logarithmic binning for $m=3$, 6, 10, 18, and 30. The solid lines are the best power-law fits. Inset: The slopes $\beta_m$ of the regression lines obtained by the ordinary linear-squares regression and the maximum likelihood estimation as a function of $m$ for $m=1, \cdots, 30$.}
\end{figure}

For the openings ${\cal{G}}_m$ with $m$ moves which appear in the $m$ layer of the Weiqi tree, we obtain the popularity $s_m$. Figure \ref{Fig:Weiqi:Open:Popularity}(a) illustrates the histograms of popularity, $H(s_m)$, for five $m$ values. $H(s_m)$ is the number of $s_m$ values in each logarithmic bin. It is trivial that openings with fewer moves can have greater popularity. We find that these histogram curves have nice power-law tails. The combination method of maximum likelihood estimation (MLE) and Kolgomonov-Smirnov test proposed in Ref.~\cite{Clauset-Shalizi-Newman-2009-SIAMR} confirms the presence of power-law tails:
\begin{equation}
  H(s_m) \propto s_m^{-\alpha_m},
  \label{Eq:Weiqi:Open:H:sm}
\end{equation}
and the tail exponents $\alpha_m$ are obtained. We also adopt the ordinary linear-squares regression (OLS) to estimate the tail exponents. For $m>10$, the first point in each curve is excluded because the first point deviates evidently from the power law. The estimated tail exponents are presented in the inset of fig.~\ref{Fig:Weiqi:Open:Popularity}(a). The OLS and the MLE methods give comparable results. It is found that the exponents $\alpha_m$ are not universal, but increases with $m$ almost linearly. This finding for Weiqi games is consistent with that for chess openings \cite{Blasius-Tonjes-2009-PRL}.

Figure \ref{Fig:Weiqi:Open:Popularity}(b) presents the complementary cumulative distributions of the relative popularity $\Pr(\geq{w_m})$ of pooled openings from ${\cal{G}}_1$ up to ${\cal{G}}_m$ for $m=3$, 6, 10, 18, and 30. We also observe nice power-law tails in the distributions with the scaling ranges spanning approximately four orders of magnitude:
\begin{equation}
  \Pr(\geq{w_m}) \propto w_m^{-\beta_m},
  \label{Eq:Weiqi:Open:CDF:wm}
\end{equation}
which is verified by the combination method of MLE and Kolgomonov-Smirnov test in Ref.~\cite{Clauset-Shalizi-Newman-2009-SIAMR}. The inset shows the dependence of the tail exponents $\beta_m$ estimated using respectively the OLS and the MLE methods on the number of moves $m$. Although the two $\beta_m$ curves do not overlap, they share very a similar tendency. The tail exponent $\beta_m$ increases with $m$ and tends to saturation:
\begin{equation}
  \lim_{m\to\infty}\beta_m = \beta,
  \label{Eq:Weiqi:beta}
\end{equation}
where $\beta\approx 1.08$ for the OLS and $\beta\approx 1.19$ for the MLE. This finding implies the possibility of universality of the tail exponent, which is again consistent with the result for chess openings \cite{Blasius-Tonjes-2009-PRL}. For chess openings, the universal exponent is reported to be ``$\alpha=2$'' \cite{Blasius-Tonjes-2009-PRL}, which is the power-law exponent of the density function, while the exponent $\beta$ in Eq.~(\ref{Eq:Weiqi:beta}) is the power-law exponent of the cumulative distribution. It follows that $\beta\approx\alpha-1$. Our results support the conjecture that: ``similar power laws could be observed in different databases and other board games, regardless of the considered game depth, constraints on player levels or the decade when the games were played'' \cite{Blasius-Tonjes-2009-PRL}.

\section{Distribution of out-degrees}

We can define the out-degree $k_{m,j}$ of any vertex $V_{m,j}$ at the $m$-th layer. As for the example Weiqi tree in fig.~\ref{Fig:Weiqi:Tree}, we have $k_{0,1}=2$ for $m=0$, $k_{1,1}=2$ and $k_{1,2}=3$ for $m=1$, and $k_{2,2}=2$, $k_{2,3}=1$ and $k_{2,5}=2$ for $m=2$. The out-degree of leaf vertices is 0 and not included in the analysis. For the real Weiqi tree, $k_{0,1}=22$ as already mentioned in the previous section. When $m=2$, the three largest out-degrees are 41, 30 and 28. When $m=29$, the maximum out-degree reduces to 3. Overall, with the increase of $m$, the maximum out-degree has a decreasing trend with mild local fluctuations.

\begin{figure}[htb]
  \centering
  \includegraphics[width=0.9\linewidth]{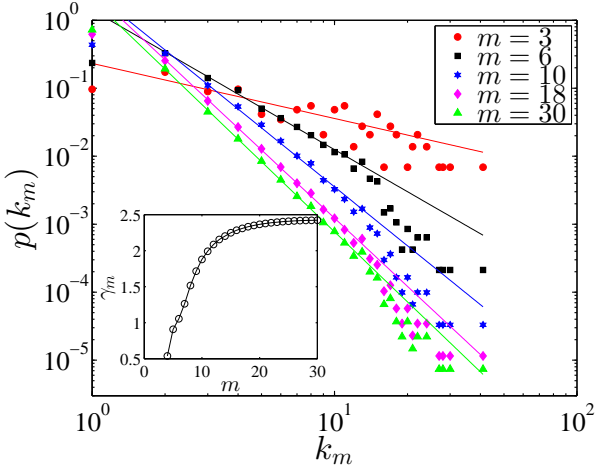}
  \caption{\label{Fig:Weiqi:Tree:Pr:k} (color online). Empirical out-degree distributions $p(k_m)$ of all vertices of opening trees ${\cal{G}}_m$ for $m=3$, $6$, $10$, $18$, and $30$. The solid lines are the best power-law fits. Inset: The exponents $\gamma_m$ of the regression lines as a function of $m$ for $m\geq3$.}
\end{figure}

We place all the out-degrees from layer $m=0$ to $m$ and estimate the underlying empirical distribution $p(k_m)$. Figure \ref{Fig:Weiqi:Tree:Pr:k} illustrates the distributions for $m=3$, $6$, $10$, $18$, and $30$. When $m$ is small, the sample is too small and the resulting distribution fluctuates a lot. The right-most point is $\max\{k_{m=1}\}=41$, which appears in all distribution curves. For all the five distributions, the bulks ($k_m>1$) exhibit nice power laws:
\begin{equation}
  p(k_m) \propto k_m^{-(\gamma_m+1)}.
  \label{Eq:Weiqi:Open:PDF:km}
\end{equation}
The inset shows that the tail exponent $\gamma_m$ increases when more layers are included in the opening tree.

\section{Distribution of branching ratios}

Now consider a vertex $V_{m,j}$ that is not a leaf. Hence, we can assume that it has daughters $V_{m+1,j_k}$ such that $V_{m,j}=\cup_kV_{m+1,j_k}$.
We calculate the branching ratios $r$ of the vertices as follows:
\begin{equation}
  r_{m,j,k}=\frac{s_{m+1,j_k}}{s_{m,j}} = \frac{s(V_{m+1,j_k})}{s(V_{m,j})}.
  \label{Eq:Weiqi:BranchingRatio}
\end{equation}
For the artificial tree in fig.~\ref{Fig:Weiqi:Tree}, there are two branching ratios $\{1/3,2/3\}$ for $m=2$, five branching ratios $\{1/3,2/3,1/3,1/6,1/2\}$ for $m=2$, and five branching ratios $\{1/2,1/2,1,1/3,2/3\}$ for $m=3$.

\begin{figure}[htb]
  \centering
  \includegraphics[width=0.9\linewidth]{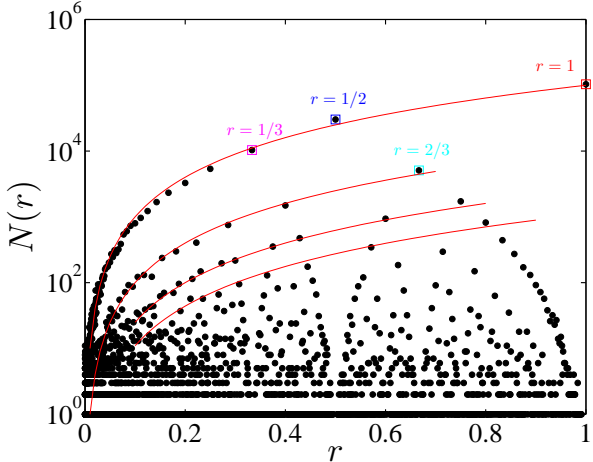}
  \caption{\label{Fig:Weiqi:Tree:N:r} (color online). Occurrence numbers of the branching ratios of non-leaf vertices of the whole Weiqi tree. The four curves are best power-law fits to four envelops which have a form of $N(r)\sim r^2$.}
\end{figure}

We calculate the branching ratios of all non-leaf vertices of the Weiqi tree. The number of distinct branching ratios is 3700. The occurrence number of each unique branching ratio is determined and presented in fig.~\ref{Fig:Weiqi:Tree:N:r}. The occurrence number of $r=1$ is 103 484, which is the most frequent. The remaining points illustrate like a half-peach and show a hierarchical pattern. The second most frequent branching ratio is $r=1/2$. Both the left and right parts with respect to $r=1/2$ look like half-peaches. The left part can be further divided into two half-peaches around $r=1/3$, while the right part can be divided into two half-peaches around $r=2/3$. This hierarchical pattern suggests an inherent self-similarity in the occurrence pattern of the branching ratios.

Intriguingly, we find that the left and right envelops of each half-peach can be well fitted by power laws:
\begin{equation}
  N(r) \propto
  \left\{
  \begin{array}{rl}
    r^{\gamma_l},  & {\mathrm{for~the~left~envelop}} \\
    -r^{\gamma_r}, & {\mathrm{for~the~right~envelop}}
  \end{array}
  \right.
\end{equation}
The exponents $\gamma_l$ and $\gamma_r$ vary for different envelops and are thus not universal. The four curves in fig.~\ref{Fig:Weiqi:Tree:N:r} are best power-law fits to four envelops with $\gamma_l=2$. These curves actually penetrate more points to the right of the original half-peaches.

\section{Distribution of uniqueness scores}

For a given game ${\cal{G}}_i$, its uniqueness score $L_i$ is the minimum number of initial moves making ${\cal{G}}_{i,m}$ different from all other openings. Speaking differently, a game's uniqueness score is the path length from the root to the corresponding leaf. For the tree in fig.~\ref{Fig:Weiqi:Tree}, $L_1=L_6=2$ and $L_2=L_3=L_7=2$, while $L_3$, $L_4$, $L_8$ and $L_9$ are longer than 3. A smaller value of uniqueness score means higher degree of uniqueness of the game.

\begin{figure}[htb]
  \centering
  \includegraphics[width=0.9\linewidth]{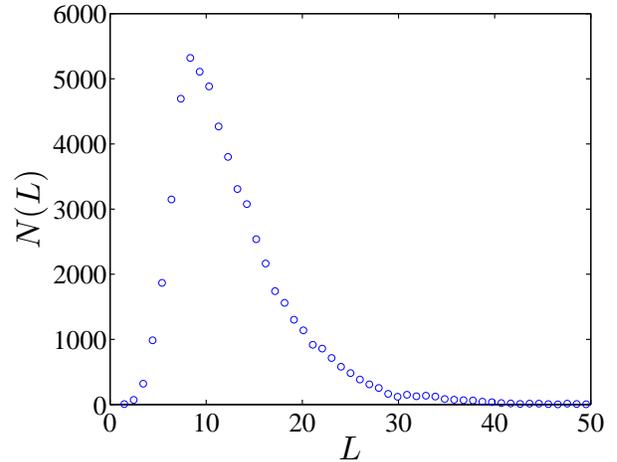}
  \caption{\label{Fig:Weiqi:Tree:N:L} (color online). Histogram of uniqueness scores of all games under investigation.}
\end{figure}

Figure~\ref{Fig:Weiqi:Tree:N:L} presents the occurrence numbers $N(L)$ of different uniqueness scores $L$. The most probable uniqueness score is $L=9$. The sample average is $\langle{L}\rangle=12.4$. The proportion of games with $5\leq{L}\leq20$ is 87.4\%. These observations are due to the fact that there are wax and wane in the popularity of Weiqi openings during different historical periods.

\section{Summary and discussions}

In summary, we have proposed to construct a directed Weiqi tree based on different sequence of moves using a database of professional Weiqi games. The construction method is the same as that for the chess tree in Ref.~\cite{Blasius-Tonjes-2009-PRL}.

We observed that the popularity distribution of Weiqi openings with a same number of moves is distributed according to a power law and the tail exponent increases with the number of moves. Moreover, the popularity distribution of Weiqi openings with the number of moves no more than a given number also have a power-law tail, and the tail exponent increases with the number of moves and seems to saturate to a universal value. These findings are robust using the ordinary least-squares regression and the maximum likelihood estimation in Ref.~\cite{Clauset-Shalizi-Newman-2009-SIAMR}. These properties are the same as for the chess tree and support the conjecture that the popularity distribution of board games follows Zipf's law with a universal exponent \cite{Blasius-Tonjes-2009-PRL}. It has been shown in Ref.~\cite{Blasius-Tonjes-2009-PRL} that the asymptotic Zipf law arises independently from the specific form of the branching ratio distribution as well as the microscopic rules of the underlying branching process. Hence the model for the chess tree applies also to the Weiqi tree.

The distribution of out-degrees is found to have a power-law form, in which the tail exponent increases with the number of moves. We also found that the distribution of branching ratios has a very complicated pattern, in which many power laws are observed. The mechanism leading to such kind of power laws is unclear and this topic is beyond the scope of the current Letter. Finally, the distribution of uniqueness scores which define the path lengths from the root vertex to the leaf vertices has a unimodal shape and most uniqueness scores are less than 30. In other words, any two arbitrary games are very unlikely to be the same after 30 moves.

An important issue concerns the impact of sample size on the observed properties. In this vein, we need to draw a clear line between possible universality and finite-size effects if larger databases are available. We are not able to significantly increase the sample size since the number of available games is indeed limited. We thus redid the analysis using 10000 randomly chosen games from the whole sample. We find that the opening popularity distribution and the complex pattern in the branching ratio distribution are not influenced by the sample size. In contrast, both the power-law tail exponent of the out-degree distribution and the average uniqueness score decrease for the smaller sample. One needs larger databases to address the question if asymptotic features exist for these two statistical properties.

Our work provides a new angle for the study of the decision making process of Weiqi playing through a directed branching tree without loops. Some of the results reported for the Weiqi tree can be interpreted by the same mechanism for the chess tree, for instance, the universal Zipf law of opening popularity. However, other empirical results such as the complicated occurrence pattern of branching ratios ask for further investigations.

\acknowledgments

We are grateful to two anonymous referees for their insightful suggestions. Errors are ours. We acknowledge financial support from the Fundamental Research Funds for the Central Universities.

\bibliography{E:/Papers/Auxiliary/Bibliography}

\end{document}